\documentclass{svjour3}                     
\smartqed  
\usepackage{graphicx}
\usepackage{amsmath}

 \journalname{Few-Body Systems}
\begin{document}

\title{Two-body problems with confining potentials
\thanks{The paper is dedicated to the 60-th birthday of Prof.\ Willibald Plessas.}
}


\author{Joseph Day \and Joseph McEwen        \and       Zoltan Papp 
}


\institute{Joseph P. Day \and Joseph E. McEwen \and Zoltan Papp \at 
              Department of Physics and Astronomy,  California State University Long Beach, Long Beach, California, USA \\
}

\date{Preprint: CSULB-PA-09-01, Received: date / Accepted: date}

\maketitle

\begin{abstract}
A formalism is presented that allows an asymptotically exact solution of non-relativistic and semi-relativistic two-body problems with infinitely rising confining potentials. We consider both linear and quadratic confinement. The additional short-range terms are expanded in a Coulomb-Sturmian basis. Such kinds of Hamiltonians are frequently used in atomic, nuclear, and particle physics.
\keywords{Relativistic quantum mechanics  \and Confining  potentials \and Meson spectra}
\end{abstract}

\section{Introduction}
\label{intro}

Non-relativistic and semi-relativistic Hamiltonians with infinitely rising potentials are often used to model quark-quark interactions or other quantum particles in a trap. 
There are many methods for solving the Schr\"odinger equation with these kinds of potentials. 
They mostly rely on variational principles and can provide satisfactory results for  the few lowest-lying states. 
However, they are unable to account for the most important property of the system; they cannot grasp the infinitely many bound states, which can be important if we want to consider three or more particles.

In this paper we offer a method that can provide an exact treatment of the infinitely rising confinement potentials. We consider both non-relativistic and semi-relativistic kinetic energy with linear and quadratic confinement. 
 We write the bound-state problem in integral equation form and incorporate   the kinetic energy and the  confining terms into the Green's operator.  Then we perform a separable expansion of the short-range terms in the Coulomb-Sturmian basis. 
 This basis allows an exact evaluation of the Green's operator in terms of continued fractions and provides  an asymptotically exact treatment of problems with linear and quadratic confinement.

This paper is organized as follows. In Sec.\ \ref{nonrel} we devise our method for non-relativistic kinetic energy with linear or quadratic confinement. Sec.\ \ref{semirel} is devoted to the semi-relativistic case. 
In Sec.\ \ref{illust} we present some numerical illustrations. We consider both linear and quadratic confinement with non-relativistic and semi-relativistic kinetic energy. Finally, in Sec.\ \ref{summary}, we summarize our results.

\section{Non-relativistic kinetic energy with short-range plus confining potential}\label{nonrel}

We consider first a Hamiltonian 
\begin{equation}\label{hnrel}
h=h^{0}+v^{(c)} +v^{(s)}~,
\end{equation}
where $h^{0}$ is the non-relativistic kinetic energy operator, $v^{(c)}$   
and $v^{(s)}$ are confining and short-range potentials, respectively.  The non-relativistic kinetic energy is given by
\begin{equation}
h^{0}=\frac{p^{2}}{2m}~.
\end{equation}
For $v^{(c)}$ we assume a polynomial potential with linear and a quadratic confinement
\begin{equation}
v^{(c)}=a_{-1}/r + a_{0}+ a_{1} r + a_{2}r^{2}~.
\end{equation}
We will see that the $1/r$ and a constant term in $v^{(c)}$ does not make the evaluation of the Green's operator any harder.

To set up an integral equation, we need the asymptotic Hamiltonian.
In this case the asymptotic Hamiltonian can be identified as
\begin{equation}
h^{c}=h^{0}+v^{(c)}~.
\end{equation}
Then  the original Hamiltonian becomes
\begin{equation}
h=h^{c}+v^{(s)}~.
\end{equation}
The bound states are the solutions of the Lippmann-Schwinger equation
\begin{equation}
|\psi\rangle=g^{c}(E)v^{(s)}|\psi\rangle~,
\end{equation}
where
\begin{equation}\label{gc}
g^{c}(z)=(z-h^{c})^{-1}~.
\end{equation}

\subsection{The Coulomb-Sturmian separable expansion method}
\label{sec4}

We solve the integral equations by using the Coulomb-Sturmian (CS) separable expansion approach.
In configuration space, the CS functions are defined by
\begin{equation}
\langle r | n l ;b\rangle = \sqrt{ \frac{ n!}{(n+2l+1)!} } (2br)^{l+1} \exp(-br) L_{n}^{2l+1}(2br)~,
\end{equation}
where $n=0,1\ldots$,  $l$ is the angular momentum, $b$ is a parameter and $L$ is the Laguerre polynomial. The biorthonormal partner to $|nl;b\rangle$ is defined as 
\begin{equation}
\langle r | \widetilde{n l;b} \rangle = 1/r  \langle r | nl;b \rangle~.
\end{equation}
The orthogonality and completeness of the basis states, in angular momentum subspace, is given by
\begin{equation}
\langle n' l';b | \widetilde{nl;b} \rangle = \langle  \widetilde{n' l';b} | n l;b\rangle =\delta_{n' n} \delta_{l' l}
\end{equation}   
and
\begin{equation}
\lim_{N\to\infty} \sum_{n=0}^{N} | \widetilde{nl;b} \rangle   \langle nl;b | = \lim_{N\to\infty} \sum_{n=0}^{N} |  nl;b \rangle  \langle \widetilde{nl;b} | = \lim_{N\to\infty} {\bf 1}^{N}_{l}= {\bf 1}_{l}~,
\end{equation}
respectively.
The CS functions also have a nice analytic form in momentum space
\begin{equation}
\langle p | n l;b\rangle = \sqrt{\frac{2}{\pi} \frac{n!}{(n+2l+1)!}} \frac{(n+l+1) l! b (4bp)^{l+1}}{(p^{2}+b^{2})^{2l+2}} G_{n}^{l+1}\left( \frac{p^{2}-b^{2}}{p^{2}+b^{2}}\right)~,
\end{equation}
and
\begin{equation}
\langle p | \widetilde{nl;b} \rangle = \frac{p^{2}+b^{2}}{2 b (n+l+1)}\langle p | n l;b\rangle~.
\end{equation}

We can approximate 
the short-range potential $v^{(s)}$ in a separable way. Here we adopt the double-basis expansion technique proposed originally in Ref.\ \cite{adhi-tomio} and extended later to Coulomb-like potentials in Ref. \cite{kdp}. It amounts to expanding $v^{(s)}$ as
\begin{eqnarray}
v^{(s)} &\approx & \sum_{n m m' n'}^{N} | \widetilde{ nl; b}  
\rangle \left( \langle  \widetilde{nl;b}| ml;b' \rangle \right)^{-1} 
\langle m l ; b' | v^{(s)} | m' l'; b' \rangle
\left( \langle   m'l';b'  | \widetilde{n'l';b} \rangle \right)^{-1} 
\langle \widetilde{ n'l'; b} |  ~ \nonumber \\
&\approx & \sum_{n n'}^{N} | \widetilde{ nl; b}  \rangle 
\underline{v}^{(s)} 
\langle \widetilde{ n'l'; b} | ~.
\label{vsep}
\end{eqnarray}
This is  an exact representation if $N$ goes to infinity and becomes an approximation if $N$ is kept finite. 
The CS matrix elements of the potential have to be evaluated numerically.  This can easily be done  in configuration or momentum space, depending how the potential is defined.

If we plug the approximated potential operator into the Lippmann-Schwinger equation we get
\begin{equation}
|\psi\rangle= \sum_{n n'}^{N} g^{c}(E)| \widetilde{ nl; b}  \rangle 
\underline{v}^{(s)} 
\langle \widetilde{ n'l'; b} | \psi\rangle~.
\end{equation}
Acting with the bra $\langle \widetilde{n'' l'';b} |$ from the left, we arrive at a matrix equation for the vector
$\underline{\psi}=\langle \widetilde{nl;b}| \psi \rangle$
\begin{equation}
\underline{\psi} = \underline{g}^{c}(E) \underline{v}^{(s)} \underline{\psi}~,
\end{equation}
where $\underline{g}^{c}(E)=\langle \widetilde{n l;b} | g^{c}(E) | \widetilde{n' l;b} \rangle $.
In fact, this equation is a homogeneous linear equation
\begin{equation}
\left( \underline{1} - \underline{g}^{c}(E) \underline{v}^{(s)} \right) \underline{\psi}=0~,
\end{equation}
whose solution can be found as zeros of the determinant
\begin{equation}
|\left(\underline{g}^{c}(E)\right)^{-1} - \underline{v}^{(s)}|=0~.
\end{equation}

\subsection{Green's operator for Hamiltonians with confinement potential} 

The Green's operator (\ref{gc}) is defined by the relation
\begin{equation}\label{jg}
J(z) \: g^{c}(z)= {\bf 1},
\end{equation}
where 
\begin{equation}J(z)=z-\frac{p^{2}}{2m} - \frac{a_{-1}}{r} -a_{1}r - a_{2}r^{2}~.
\end{equation}
This equation in the CS basis looks like
\begin{equation}
\sum_{n'} \langle nl;b | J(z) | n'l;b \rangle \; \langle \widetilde{n'l;b } | g^{c}(z) |\widetilde{ n'' l;b  } \rangle= \delta_{n n''}~,
\end{equation}
and we need the $N\times N$ upper left part of the $\infty\times \infty$ Green's matrix.

The operator $J$ has an infinite matrix representation. It 
can be constructed from the following matrix elements \cite{kelbert}:
\begin{equation}
\langle{nl;b}|\frac{1}{r} |{n'l;b}\rangle=\langle{n'l;b} | \frac{1}{r} |{nl;b}\rangle=\delta_{n n'}
\end{equation}
\begin{equation} 
\langle{nl;b}|{n'l;b}\rangle=\langle{n'l;b}|{nl;b}\rangle=\begin{cases} 
\frac{1}{b}(n+l+1)&\text{for  } n'=n\\ 
-\frac{1}{2b}\sqrt{n'(n'+2l+1)}&\text{for  } n' =n+1\\
0 &\text{for   } n'> n+1
\end{cases} 
\end{equation}
\begin{equation}
\langle{n'l;b} | \frac{p^{2}}{2m} | {nl;b}  \rangle= \langle{nl;b} |\frac{p^{2}}{2m} |{n'l;b}\rangle=\begin{cases} 
\frac{b}{2m}(n+l+1)&\text{for  } n'=n\\ 
-\frac{b}{4m}\sqrt{n'(n'+2l+1)}&\text{for  } n' =n+1\\
0 &\text{for   } n'> n+1
\end{cases} 
\end{equation}

\begin{equation}
\begin{split}
\langle{n'l;b}| r|{nl;b} \rangle& =\langle{nl;b}| r | {n'l;b} \rangle= \\
& = \begin{cases} 
\frac{b}{2}(n+l+1)&\text{for  } n'=n\\ 
-\frac{b}{4}\sqrt{n'(n'+2l+1)}&\text{for  } n' =n+1\\
\frac{1}{4b^2} \sqrt{n'(n'-1)(n'+2l)(n'+2l+1)} &\text{for   } n'= n+2\\
0 &\text{for   } n'> n+2
\end{cases} 
\end{split}
\end{equation}

\begin{equation}
\begin{split}
& \langle{n'l;b} |r^2 | {nl;b} \rangle =\langle{nl;b} |r^2 | {n'l;b} \rangle=\\
&=\begin{cases} 
-\frac{1}{8b^3} [(((10n+2l+4)(n+2l+3) +\\
\hspace{2cm} 9n(n-1))(n+2l+2)+n(n-1)(n-2))]   &\text{for  } n'=n\\ 
-\frac{3}{8b^3}[(4n'+2l)(n'+2l+2)+(n'-1)(n-2) ]\sqrt{n'(n'+2l+1)} 
&\text{for  } n' =n+1\\
\frac{3}{8b^2} \sqrt{n'(n'-1)(n'+2l)(n'+2l+1)} 
&\text{for   } n'= n+2\\
- \frac{1}{8b^3}(2n'+2l)\sqrt{n'(n'-1)(n'-2)(n'+2l+1)(n'+2l)} 
&\text{for   } n'= n+3\\
0 &\text{for   } n'> n+3
\end{cases}
\end{split}
\end{equation}

In our previous study we found that the inverse of an infinite symmetric tridiagonal  matrix can be determined by a continued fraction \cite{klp}. 
In this case, if the $r^{2}$ term is present in $J$, its matrix representation is a septadiagonal infinite symmetric band matrix. Such a matrix can be considered as a tridiagonal matrix of $3\times3$ block matrices.  
If $r$ is the highest power in $J$, the matrix is pentadiagonal, which can be considered as tridiagonal matrix of $2\times2$ blocks. An infinite symmetric band matrix can be considered as a tridiagonal matrix of block matrices. The  corresponding Green's matrix can be constructed by matrix continued fractions \cite{kelbert}:
\begin{equation}
\underline{g}^{c}(z) = (\underline{J} - \delta_{i, N'} \delta_{j, N'} J_{N',N'+1} C_{N'+1} 
J_{N'+1,N'})^{-1},
\end{equation}
where $C$ is a matrix continued fraction
defined recursively by
\begin{equation}
C_{i+1}=(J_{i+1,i+1}-J_{i+1,i+2} C_{i+2} J_{i+2,i+1})^{-1}~
\end{equation}
and $J_{i,j}$ refers to $3\times3$ or $2\times2$ blocks. The index $N'=N/3$ or $N'=N/2$ depending on whether we cover $\underline{J}$ by $3\times3$ or $2\times2$ blocks.
We found, that $\underline{g}^{c}$ is an inverse of the modified $\underline{J}$ matrix. The modification is a matrix continued fraction $C$, which only effects the bottom right $3\times 3$ or $2\times2$ block of $\underline{J}$.

\section{Semi-relativistic kinetic energy with short-range  plus confining potential}\label{semirel}

Now we consider the problem of the previous section with semi-relativistic kinetic energy
\begin{equation}
h=h^{0}_{r}+v^{(c)} +v^{(s)}~,
\end{equation}
where 
\begin{equation}\label{relkin}
h^{0}_{r}=\sqrt{m^{2}+p^{2}}-m~.
\end{equation}
The semi-relativistic kinetic energy operator in the CS basis does not have a tridiagonal or band-matrix structure. Numerical studies show that its matrix representation for large $n$ and $n'$ indices is pentadiagonal dominant (see Table \ref{tab0}). These asymptotically dominant elements can easily be taken into account  in our matrix continued fraction method formalism.

\begin{table}[h]
\caption{The CS matrix elements of $h^{0}_{r}=\sqrt{m^{2}+p^{2}}-m$ for $n,n'=170-179$. The penta-diaognal elements are clearly dominant. We used $m=0$, $l=0$, and $b=4$. }
\label{tab0}       
\begin{tabular}{rrrrrrrrrr}
\hline
    80.   &   10.   &   -30.   &    -4.    &   -5.   &    -2.    &   -2.    &   -1.   &    -1.  &     -1. \\
      10.   &    81.   &    10.   &   -30.   &    -4.    &   -5.    &   -2.    &   -2.    &   -1.   &    -1. \\
     -30.    &   10.   &    81.    &   10.   &   -30.   &    -4.    &   -5.    &   -2.   &    -2.   &    -1. \\
      -4.   &   -30.    &   10.      &      82.    &  10.   &   -30.    &   -4.   &    -5.   &    -2.    &   -2. \\
      -5.    &  -4.   &   -30.   &   10.    &  82.  &    10.  &   -31.   &   -4.   &   -5.   &   -2. \\
      -2.    &  -5.   &   -4.   &  -30.   &   10.   &   82.   &   10.  &   -31.   &   -4.   &   -5.\\
      -2.    &  -2.   &   -5.   &   -4.   &  -31.    &  10.   &   83.   &   10.  &   -31.   &   -4.\\
      -1.    &  -2.   &   -2.   &   -5.   &   -4.  &   -31.   &   10.   &   83.   &   10.   &  -31.\\
      -1.    &  -1.   &   -2.   &   -2.   &   -5.   &   -4.   &  -31.   &   10.  &    84.  &    10.\\
      -1.   &   -1.  &    -1.   &   -2.  &    -2.   &   -5.  &    -4.   &  -31.   &   10.   &   84. \\
\hline
\end{tabular}
\end{table}

\section{Numerical illustrations}\label{illust}

\subsection{Yukawa plus linear confinement}

\begin{figure}
  \includegraphics[width=8cm]{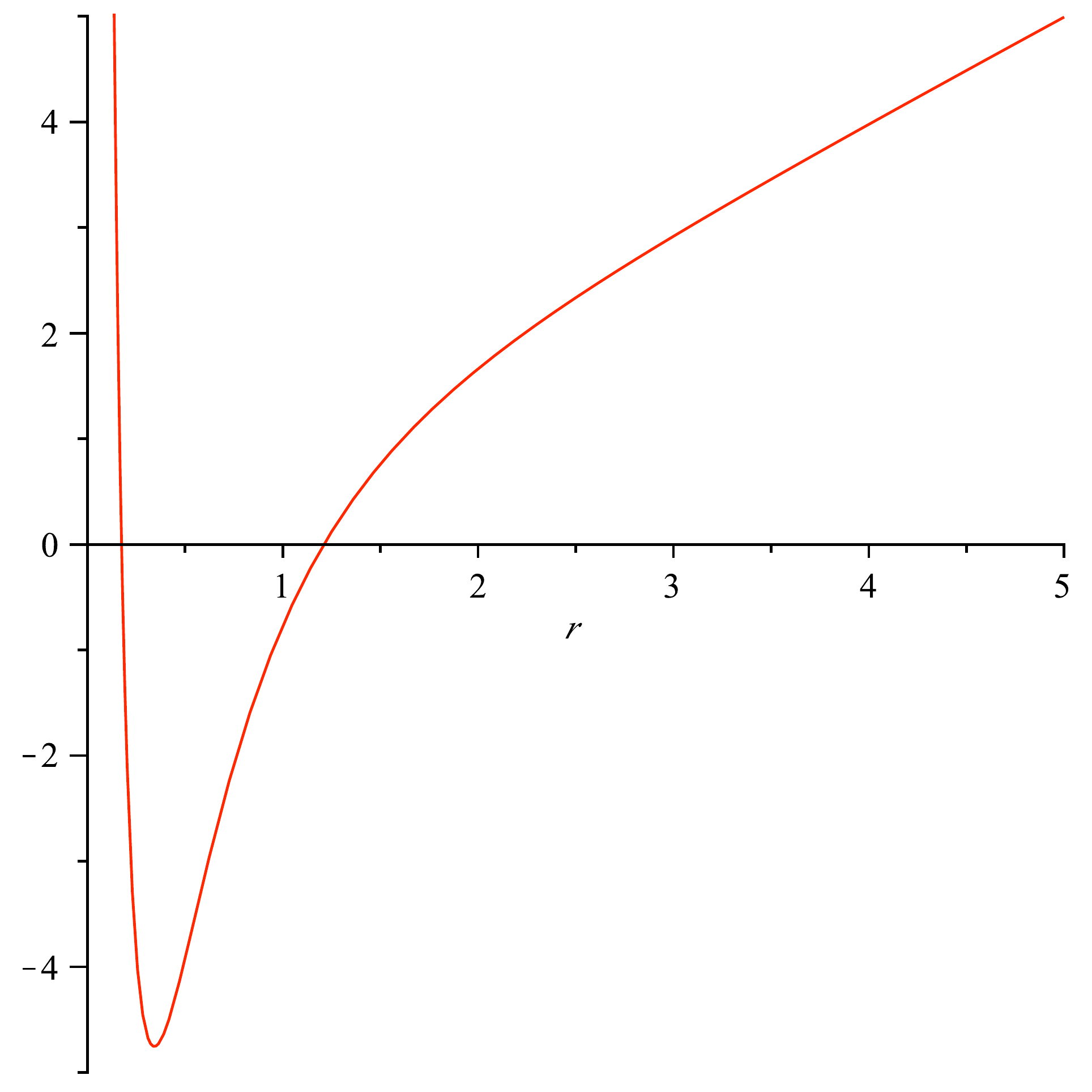}
\caption{Yukawa plus linear potential $v(r)=10\exp(-5r)/r-5\exp(-r)/r+r$.}
\label{fig:1}       
\end{figure}

We illustrate the efficiency of our method with the example of non-relativistic kinetic energy with Yukawa plus linear confinement potential,
\begin{equation}
v(r)=a_{1}r + v_{1}\exp(-\mu_{1} r )/r + v_{2}\exp(-\mu_{2} r )/r .
\end{equation}
We choose the parameters as $m=1$, $a_{1}=1$, $v_{1}=10$, $\mu_{1}=5$,  $v_{2}=-5$, and $\mu_{2}=1$. This potential is shown in Fig.\ \ref{fig:1}. Similar types of potentials are used to model the quark-quark interaction. 
We found that the parameters for the basis of the separable expansion $b=4$ and $b'=3$ were optimal.
  
\begin{table}[h]
\caption{The first three eigenstates in a Yukawa plus linear confinement in terms of basis states. }
\label{tab:1}       

\begin{tabular}{c| lll | c | lll  }
\hline\noalign{\smallskip}
 &\multicolumn{3}{c}{Non-Relativistic}   &  \multicolumn{3}{r}{Semi-Relativistic} \\
$N$ & 1st & 2nd & 3rd & $N$ & 1st & 2nd & 3rd  \\
\noalign{\smallskip}\hline\noalign{\smallskip}
$6$  &   -0.4380704  &  2.396969   &  3.820619 & $18$ & -1.596369 &   1.007886   &     2.324412  \\
$9$ &    -0.4380688  &  2.399815   &  3.839923   & $21$ & -1.596349 &   1.007905   &     2.324431 \\
$12$ &   -0.4380692 &  2.399815 &    3.840087  & $24$ & -1.596354 &   1.007884   &     2.324392 \\
$15$ &   -0.4380692 &  2.399815  &   3.840087  & $27$ & -1.596355 &   1.007883   &     2.324391 \\ 
$18$ &  -0.4380692 &   2.399815  &   3.840087   & $30$ & -1.596355 &   1.007884   &     2.324394 \\
$21$ &   -0.4380692 &  2.399815  &   3.840087 & $33$ & -1.596355 &   1.007884   &     2.324394  \\
\noalign{\smallskip}\hline
\end{tabular}
\end{table}

 What is covered by $2\times 2$ matrices is also covered by $3\times 3$ matrices, therefore we use $3\times 3$ blocks throughout. For a direct comparison of relativistic effects, we  consider the non-relativistic and semi-relativistic Hamiltonian with the same potential.
The convergence of the first three eigenstates of the non-relativistic and semi-relativistic Hamiltonian with increasing $N$ are shown in Table \ref{tab:1}. To achieve a similar accuracy we need more basis states in the semi-relativistic case, but in both cases excellent accuracy has been observed by using a relatively small basis size.

\subsection{Yukawa plus quadratic confinement}

\begin{figure}[h]
  \includegraphics[width=8cm]{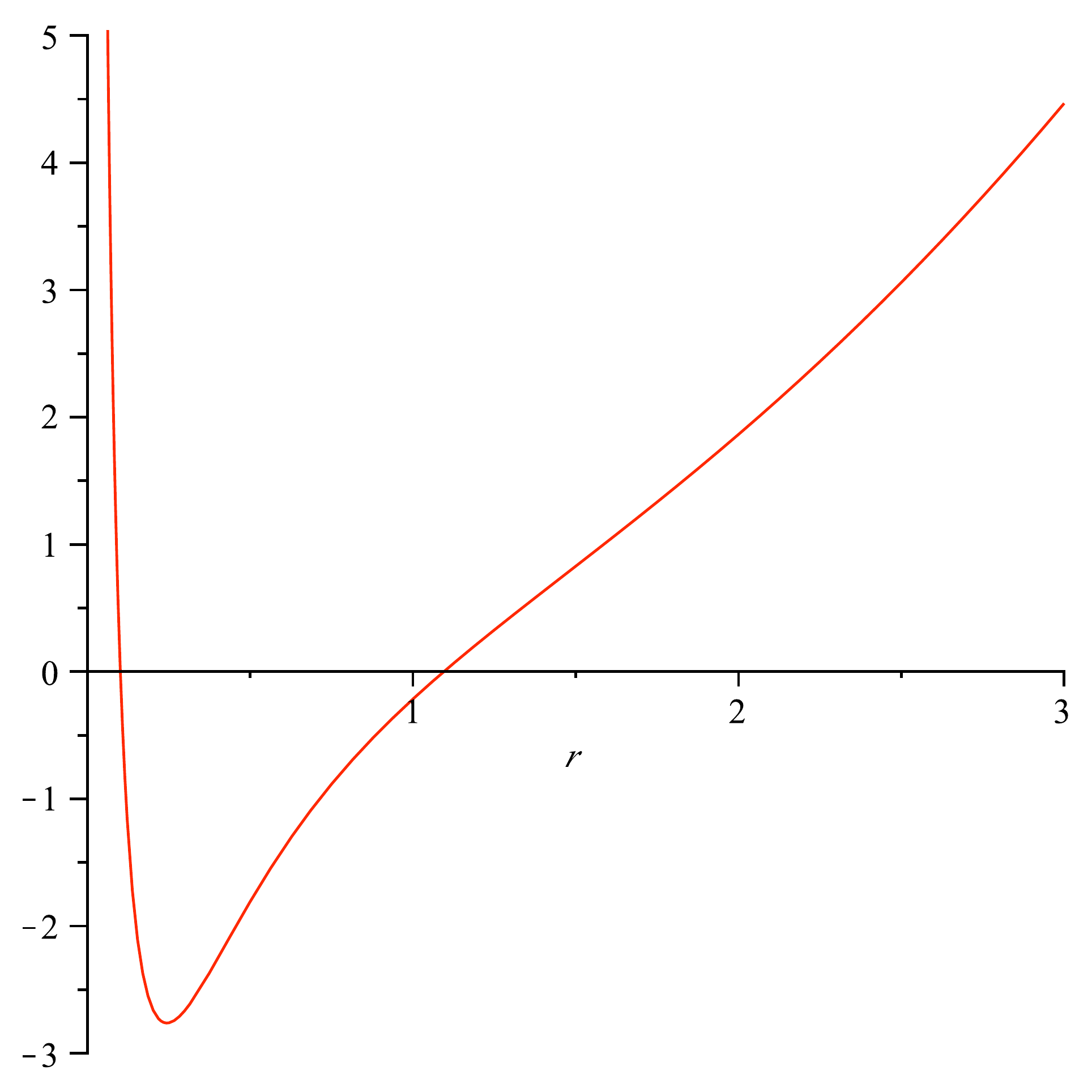}
\caption{Yukawa plus quadratic potential $v(r)=3\exp(-5r)/r-2\exp(-r)/r+0.5 r^{2}$.}
\label{fig:2}       
\end{figure}

Another example we consider is Yukawa plus quadratic confinement potential (Fig.\ \ref{fig:2})
\begin{equation}
v(r)=a_{2}r^{2}+ v_{1}\exp(-\mu_{1} r )/r + v_{2}\exp(-\mu_{2} r )/r  ~,
\end{equation}
with  a non-relativistic and semi-relativistic kinetic energy operator.
We use the parameters $m=1$, $a_{2}=0.5$, $v_{1}=3$, $\mu_{1}=5$, $v_{2}=-2$, and $\mu_{2}=1$ .

The convergence of the first three eigenstates of the non-relativistic and semi-relativistic Hamiltonian with increasing $N$ are shown in Table \ref{tab:2}. We can draw the same conclusion as before; excellent accuracy can be achieved using a 
a relatively small basis, but  the convergence is not as rapid in the semi-relativistic case.

\begin{table}[h]
\caption{The first three eigenstates of a Yukawa plus quadratic confining potential in terms of basis states. }
\label{tab:2}       

\begin{tabular}{c| lll  | c | lll  }
\hline\noalign{\smallskip}
& \multicolumn{3}{c}{Non-Relativistic}   &  \multicolumn{3}{r}{Semi-Relativistic} \\
$N$ & 1st & 2nd & 3rd & $N$ & 1st & 2nd & 3rd  \\
\noalign{\smallskip}\hline\noalign{\smallskip}
$6$  &    0.5502992   &   2.929190   &   5.040812 & $18$ & -0.1080278  &    1.603202   &     2.854743  \\
$9$ &     0.5503106   &   2.929148   &   5.056162 &$21$ &  -0.1080290 &     1.603201  &      2.854741   \\
$12$ &   0.5503106   &   2.929150  &    5.056224   & $24$ &  -0.1080297 &    1.603200    &       2.854741\\
$15$ &   0.5503106   &   2.929150  &    5.056225  & $27$ &  -0.1080301 &    1.603200   &       2.854740 \\ 
$18$ &   0.5503106   &   2.929150  &    5.056225   & $30$ &  -0.1080304 &    1.603200   &       2.854740 \\
$21$ &   0.5503106   &   2.929150  &    5.056225  & $33$ &  -0.1080306 &    1.603199   &        2.854740 \\
\noalign{\smallskip}\hline
\end{tabular}
\end{table}

\section{Summary and conclusions}\label{summary}

We developed a method for solving two-body non-relativistic and semi-relativistic problems 
with confining potentials. We considered both linear and quadratic confinements. This method is based on a separable expansion of the short-range interaction in terms of CS basis and
 involves the evaluation of the corresponding Green's operator on that basis. The CS basis is particularly advantageous as it has simple analytic forms in configuration and momentum spaces allowing an easy evaluation of the potential and kinetic energy matrix elements. 
The non-relativistic kinetic energy plus confinement potential is represented by a infinite symmetric band matrix. The evaluation of the Green's operator amounts to inverting the infinite band matrix 
by means of matrix continued fractions.  We found numerically that the semi-relativistic kinetic energy operator is asymptotically pentadiagonal in the CS basis.  Therefore the Green's matrix of the semi-relativistic problem can also be evaluated using matrix continued fractions. As numerical examples we took the Yukawa plus linear and quadratic confinement models. We achieved very accurate results using a small number of basis functions.

\section{Acknowledgement}    
 Z.~P.\ is thankful to Prof.\ Plessas for twenty years of friendship.


\end{document}